\documentclass[letterpaper, 10 pt, conference]{ieeeconf}  % Comment this line out if you need a4paper

\IEEEoverridecommandlockouts                              % This command is only needed if 
\usepackage[T1]{fontenc}
\usepackage{flushend}
\usepackage{pgf,tikz}
\usepackage{mathrsfs}
\usetikzlibrary{arrows}
\usepackage{amsmath}
\usepackage{algorithm}
\usepackage[noend]{algpseudocode}
\usepackage{tikz}
\usepackage[export]{adjustbox}
\usetikzlibrary{patterns,arrows,calc,decorations.pathmorphing}
\definecolor{ududff}{rgb}{0.30196078431372547,0.30196078431372547,1}

\title{\LARGE \bf
Terrain-based vehicle localization using an active suspension system}

\author{Yu Jiang$^{1,2}$, John Eisenmann$^{1}$, William Graves$^{1}$, Vijayaraghavan Sridhar$^{1,3}$, and Zackary Anderson$^{1}$ % <-this % stops a space}
\thanks{$^{1}$ Y. Jiang, J. Eisenmann, W. Graves, and Z. Anderson are with the CTO Office at ClearMotion, Inc., 805 Middlesex Turnpike, Billerica, MA 01821, USA}
\thanks{$^{2}$ Corresponding author: Y. Jiang, \tt\small yjiang@clearmotion.com}%
\thanks{$^{3}$ The work described here was completed while V. Sridhar was with ClearMotion{}}%
}

\begin{document}

\maketitle
\thispagestyle{empty}
\pagestyle{empty}

%%%%%%%%%%%%%%%%%%%%%%%%%%%%%%%%%%%%%%%%%%%%%%%%%%%%%%%%%%%%%%%%%%%%%%%%%%%%%%%%
\begin{abstract}

This paper, for the first time, presents a terrain-based localization approach using sensor data from an active suspension system. The contribution is four-fold.
First, it is shown that a location dependent road height profile can be created from sensor data of the active suspension system. 
Second, an algorithm is developed to extract a pitch profile from the road height profile data. The ideal pitch profile is vehicle-independent and only depends on the road. This pitch profile  generated from an on-board computer is matched with a known terrain map to achieve real-time positioning.
Third, a crowd-sourced map creation algorithm is developed to create and improve the terrain map that contains pitch profile.
Fourth, experiments have been conducted to validate the accuracy and robustness of the proposed localization approach. 

\end{abstract}

%%%%%%%%%%%%%%%%%%%%%%%%%%%%%%%%%%%%%%%%%%%%%%%%%%%%%%%%%%%%%%%%%%%%%%%%%%%%%%%%
\section{INTRODUCTION}
As a critical capability for autonomous and intelligent vehicles, localization makes it possible to determining precise position of the vehicle on the map. The typically used tool for outdoor localization is GPS, which is limited by low rate, accuracy, and satellite accessibility. An alternative approach is vision-based localization (VBL), which has become an active research area in the past decade (see, for example, the survey paper \cite{piasco2018survey} and the work \cite{karlsson2005vslam} on vSLAM). VBL usually requires higher computing power for video and/or LiDAR point-cloud processing in real-time. It also requires an high definition (HD) map that contains detailed features in the environments. Probabilistic models are often employed to handle the uncertainty in the change of landmarks, as well as sensor noise.

Terrain-based localization (TBL) is referred to as matching in-vehicle sensor measurements with a stored topographical terrain map, and taking the location of the best match to be the position of the navigator. As the primary localization method before GPS was invented, terrain-based localization has been used extensively in aircraft \cite{gustafsson2002particle}, cruise missile applications \cite{hostetler1978optimal}, and underwater systems \cite{whitcomb2000advances}.

For road vehicles, map-based localization with particle filters were proposed in \cite{gustafsson2002particle}, in which the driven path was matched to a road map. However, driven path may not contain enough significant features on a small scale, and it can be easily affected by noise, such as traffic volume, it along does not always provide enough localization accuracy. In 2001, it was first pointed out in \cite{bae2001road} that low frequency component of pitch measurement correlates well with road grade, since the road grade usually changes slower than the pitch motion of the vehicle. Pitch profile was not reportedly used for localization until 2006, when location dependency and repeat-ability of the pitch profile in spatial domain were demonstrated \cite{martini2006gps}, and offline longitudinal positioning was studied by searching the best correlation between a new pitch profile, reported by in-vehicle inertial measurement units (IMUs) devices,  and the known accurately measured pitch profile. Particle filters were later used to match the pitch measurements onto the terrain map to achieve real-time localization \cite{dean2011terrain}. It has also been shown that pitch profile can be used in combination with driven path to create a hybrid map for improved localization accuracy \cite{li2014extended}. To overcome the potential zero drift in IMU measurements and to enhance robustness, pitch difference, instead of pitch profile itself, was used for matching with
Bayesian inference and particle filters \cite{li2016robust}. In \cite{laftchiev2015vehicle}, pitch profile data was encoded using linear models for computational efficiency. 

To the best of our knowledge, in all the above-mentioned methods, pitch profiles are measured using in-vehicle IMUs.  As a result, these approaches usually ignore the unmodeled dynamics within the suspension systems. Since the suspension dynamics can vary greatly from one vehicle to another, Pitch profile created by one vehicle may not be applicable to the other. In addition, with the development of active and semi-active suspension control technologies, the suspension dynamics of the same vehicle may vary drastically, causing potential inconsistency in the estimated pitch profiles. Although the previous methods all used low-pass filters to get rid of the noise caused by unmodeled dynamics to increase consistency, they discarded quite a few useful statistics in the data. This could affect localization accuracy.

In this paper, we propose a new methodology to conduct longitudinal localization using sensors from an active suspension control system. An active suspension control system requires high-quality and high-frequency sensors (e.g., wheel accelerometers) mounted in the wheel assembly. With the help of these sensors, it is practical to reconstruct cleaner and more repeatable pitch profiles, with comparably higher resolution than traditional methods. In addition, we propose a dynamic matching and merging algorithm to create and maintain a crowd-sourced terrain map. The terrain map has sufficient information to generate a {\it master} pitch profile that is compared with the {\it live} pitch profile as generated from in-vehicle computers. Finally, we present our experimental results to validate the proposed methodology.

The remainder of paper is organized as follows. Section II reviews a classical quarter-car model with active suspension control systems, and provides several approaches to reconstruct repeatable road profile. Section III details the main results of the proposed technology, including how the pitch profile is generated, and how the crowd-sourced map is created and updated. Section IV presents the experiments we conducted and related data analysis. Section V gives conclusions and briefly points out future research directions. 

\section{PRELIMINARIES}
\label{roadprofile}
\subsection{Road profile reconstruction}

Consider a quarter car suspension system \index{Car suspension system} described by the following ordinary differential equations \cite{hrovat1993applications}:
\begin{eqnarray}
\dot x_1 &=& x_2 \label{eq:qcar:1}\\
\dot x_2 &=& -\frac{k_ss_d+b_s(s_v)-f}{m_b}\label{eq:qcar:2}\\
\dot x_3 &=& x_4\label{eq:qcar:3}\\
\dot x_4 &=& \frac{k_ss_d+b_s(s_v)-k_t(x_3-r)-f}{m_w}\label{eq:qcar:4}
\end{eqnarray}
where $x_1$, $x_2$, and $m_b$ denote respectively the position, velocity, and mass of the quarter car body;  $x_3$, $x_4$, and $m_w$ represent respectively the position, velocity, and mass of the  wheel assembly; $s_d=:x_1-x_3$ and $s_v:=x_2-x_4$ are the displacement and velocity of the shock absorber; $k_t$, $k_s$, are the tire and the spring coefficients; $b_s(\cdot)$ is a nonlinear damping function; $f$ is the force applied between the body and the wheel assembly.

Our road profile estimation method is based on the quarter car model and will be elaborated in the next Section.

\subsection{Cross-correlation}

This paper employs cross-correlation \cite{buck1997computer} \cite{stoica2005spectral} to find a signal snippet within a longer stream. The theoretical cross-correlation sequence of two jointly stationary random process $x[n]$ and $y[n]$ is given by
\begin{eqnarray}
R_{xy}[m] &=& E\{x[n+m]y^*[n]\} \\
&=& E\{x[n]y^*[n-m]\} 
\end{eqnarray}
where $n\in(-\infty,\infty)$, asterisk denotes complex conjugation, and $E(\cdot)$ stands for the expected value operator.

In practice, only a finite segment of the random process is available. Thus, the raw correlation without normalization is defined as
\begin{eqnarray}
\hat{R}_{xy}[m] = \left\{\begin{array}{cc}
\sum\limits_{n=0}^{N-m-1}x[n+m]y[n]^*,& m\ge 0,\\
\hat{R}^*_{xy}[-m], & m<0.
\end{array} \right.
\end{eqnarray}
When applying raw cross-correlation on a signal snippet and a longer stream, the point at which the cross-correlation is the greatest may indicate the delay between the starting points. Thus, it is possible to apply cross-correlation between a live profile generated by the on-board computer, and a master profile stream stored in the cloud to localize the vehicle. This will be detailed in Section \ref{sec:loc}.

\section{TERRAIN-BASED LOCALIZATION}
This section presents the main results of the localization methodology, which includes pitch profile creation, map merging, online localization, and some key implementation issues.  
\label{sec:loc}

\subsection{Road profile estimation}
On distinctive advantage of using an active suspension system for this task is that we can conveniently identify the parameters of the quarter car model. Indeed, linearizing the damping term $b_s(s_v)$ to $b_s s_v$ makes the quarter car model \eqref{eq:qcar:1}-\eqref{eq:qcar:4} a standard LTI system, and quite a few system identification approaches can be directly applied. Notice that this is not straightforward in a passive or semi-active system, due to the lack of sensors and the ability to inject actuation force.

With identified parameters of the quarter car model \eqref{eq:qcar:1}-\eqref{eq:qcar:4}, the road profile at each corner of the vehicle is estimated as 
\begin{eqnarray}
\hat{r}=\frac{1}{k_t}({m_w\dot{x}_4 -k_s s_d-b_s(s_v)+f}) + \hat{x}_3
\end{eqnarray}
where $\dot{x}_4$ is directly reported from the wheel accelerator, and $\hat{x}_3$ is computed by integrating $\dot{x}_4$ twice and applying a high-pass filter to remove the low-frequency numerical drift caused by integration. 

For other approaches, such as observer-based profile reconstruction methods, please refer to \cite{noll2016accuracy} and \cite{doumiati2014adaptive} as examples.

\subsection{Pitch profile creation}

The road profile $r(t)$ described in Section \ref{roadprofile} is a function of time. With standard vehicle velocity measurement (e.g., 20Hz to 50Hz collected from CAN bus), the road profile can be converted to a function of distance ${r}_d(s)$ based on the following equality:
\begin{eqnarray}
{r}_d\left(\int_0^tv(\tau)d\tau\right) = {r}(t)
\end{eqnarray}
where $v(t)$ is the vehicle speed. 

A detailed conversion algorithm is given in Algorithm \ref{alg:profileconv}, where $d_{map}$ is the discrete point on the map, and $\Delta d$ is the resolution of the map.
\begin{algorithm}
\caption{Road profile conversion}\label{alg:profileconv}
\begin{algorithmic}[1]
\Procedure{Initialization:}{}
\State $d_{old} \gets 0$
\State $d_{new} \gets 0$
\State $t_{old} \gets 0$
\State $r_{old} \gets 0$
\State $r_d \gets \emptyset$
\State $d_{map} \gets 0$
% \State $\textit{mappoint}_{next} \gets \Delta d$
\EndProcedure
\Procedure{Time to distance conversion}{}
\While {receiving new data $(t, v, r)$ and $v>0$}
\State $d_{old} \gets d_{new}$
\State $d_{new} \gets d_{new} + \int_{t_{old}}^t~v~dt$
\State $t_{old} \gets t$
\If {$d_{map}\in(d_{old}, d_{new})$}
\State $r_{d, new} \gets \frac{(r-r_{old})(d_{map}-d_{old})}{d_{new} - d_{old}}+r_{old}$

\State Append $r_{d, new}$ to the end of $r_d$
\State $d_{map}\gets d_{map} + \Delta d$
\EndIf
\EndWhile
\State return $r_d$
\EndProcedure
\end{algorithmic}
\end{algorithm}
\noindent 

The quarter car model associated with each wheel can generate such a road profile as a function in distance. Thus, for a regular four-wheel vehicle, four different road profiles are generated in the same way.

It can be converted to the pitch profile of the chassis by 
\begin{eqnarray}
p(s) = \tan^{-1}\left(\frac{r_{fl}(s)+r_{fr}(s)-r_{rl}(s)-r_{rr}(s)}{2l_{b}}\right)\nonumber\\
&&
\end{eqnarray}
where $l_b$ is the wheelbase (the distance between the front and rear axles of a vehicle), $r_{fl}$, $r_{fr}$, $r_{rl}$, and $r_{rr}$ are the road profile estimated at front-left, front-right, rear-left, and rear-right wheels, respectively.

In order to eliminate the drift caused by integrating of wheel accelerations, we differentiate the pitch profile $p(s)$ with respect to distance to extract the most consistent portion of the data across different drives.

Note that the pitch profile depends on the vehicle wheelbase, which may vary from one car to another. Therefore, the terrain map only stores the road height profile but will derive the pitch profile whenever it is needed.

% Crowd sourced mapping
\subsection{Crowd-sourced map creation}
Road profile height and the resultant pitch profile estimated in this way are by no means to serve only one car in a specific area. Instead, it is desired to collect the profiles generated from different vehicles and attach the data to a database of geographical information, e.g., OpenStreetMap (OSM) \cite{haklay2008openstreetmap}, such that all vehicles having access to the database and with sufficient sensors can share the profile data and use it for its own localization. To this end, we have developed a new approach for creating and continuously improving a crowd-sourced terrain map that contains road profiles.

\begin{figure}
%\hspace{-2mm}
\centering
\includegraphics[width = 0.55\textwidth]{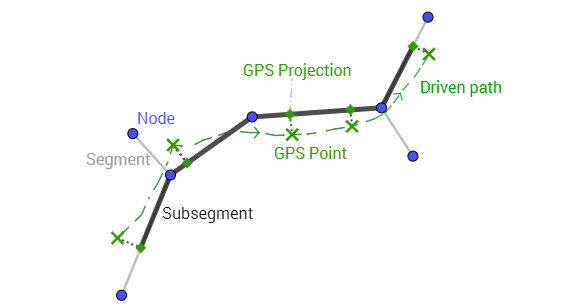}
\caption{Graph representation of a map. Each circle filed with blue color is a {\it node}, which consists of a single point in space defined by its latitude, longitude and node id. A straight section of a way between exactly two adjacent nodes is called as a {\it segment}. Each green cross is a {\it GPS point} measured by the on-board GPS device. Each GPS point has a projection on its nearest segment. A straight section between a GPS projection point and a node is referred to as a {\it subsegment}.}
\label{fig:graphmap}
\end{figure}

To begin with, given a graph representation (Fig. \ref{fig:graphmap}) of the map geometry, one can define its associated road profile data with a desired map resolution. For example, one can assign road height values every $0.1m$ along a segment. These values are stored in the cloud and the values from adjacent segments can be loaded and stitched together to create a {\it master} profile. On the other hand, a vehicle operated in the same area may generate a {\it live} profile with its on-board computer and send the data to the cloud for further improvement of the existing master profile.

\begin{figure}[!ht]
\centering
%\hspace{-2mm}
\includegraphics[width = 0.55\textwidth]{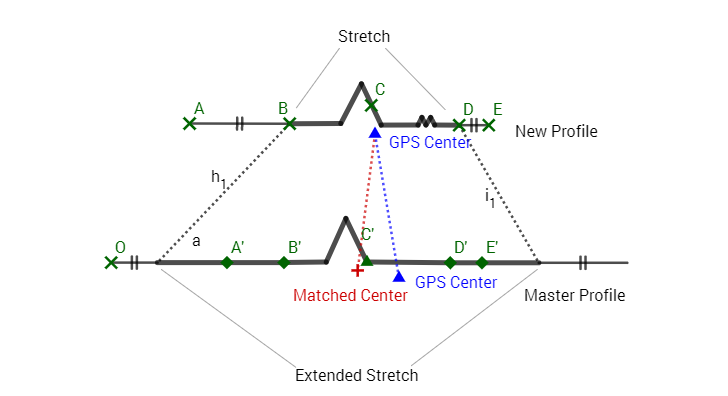}
\caption{Matching a live stretch with an extended stretch in the master profile. The extended stretch is found via the GPS center of the live stretch. The key in the matching algorithm is to precisely align the center of the live stretch with the corresponding location in the extended stretch. This location may be different from the location with the same GPS coordinate, due to the inaccuracy in GPS measurements.}
\label{fig:mapStretch}
\end{figure}

\begin{figure}[!ht]
\centering
\includegraphics[width = 0.5\textwidth]{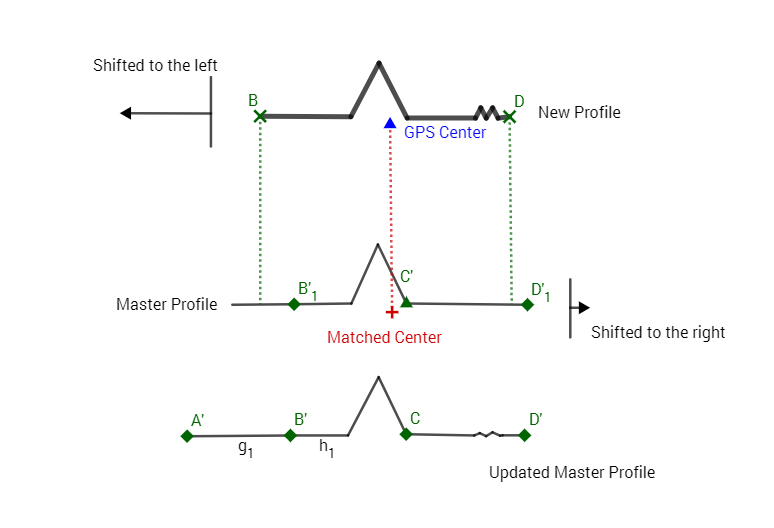}
\caption{Once the center of the live stretch is matched with a location point in the extended stretch, one can crop the extended profile to obtain a portion of the master profile that has the same length as the stretch. Then, update the master profile by averaging it with the stretch. The GPS data of the master profile may also be updated (in terms of shifting the profile towards the left or the right) with the GPS data of the stretch.}
\label{fig:mapAverage}
\end{figure}

Once a live profile is received, the associated driven path is matched onto the map by using the on-board GPS measurements (Fig. \ref{fig:graphmap}). Next, the live profile is broken down into into multiple {\it stretches}. A stretch is defined as a piece of road profile with approximately $100m$ long, and it is associated with a pair of GPS points for its start and end points. The stretch has a center GPS point, using which we can find an extended stretch, which is a longer piece in the master profile that has the same GPS center as the stretch (Fig. \ref{fig:mapStretch}). Then, we try to find the precise match of the stretch within the extended stretch. This can be done by performing cross-correlation between the related pitch profiles. If this is successful, we merge the live stretch into the master profile by doing a weighted average (Fig. \ref{fig:mapAverage}). This merge operation will not only improve the height profile but also improve the GPS coordinates of the master profile.

The map creation and improvement algorithm is summarized in Algorithm \ref{alg:mapcreation} and illustrated in Fig. \ref{fig:mapStretch} and Fig. \ref{fig:mapAverage}.

\begin{algorithm}
\caption{Terrain map creation and improvement}
\label{alg:mapcreation}
\begin{enumerate}
\item Initialize the master profile with zeros.
\item Receive a live height profile as a function of distance from a vehicle.
\item Take a stretch from the live profile.
\item Find the GPS coordinate of the stretch center.
Get an extended stretch from the master profile with the same GPS coordinate.
\item Apply raw cross-correlation to precisely match the stretch within the extended stretch.
\item Crop the extended profile to get the piece of master profile that matches with the stretch.
\item Merge the stretch into the master profile.
\item If there are new available stretches to use, go to 3).
\end{enumerate}
\end{algorithm}

\subsection{Real-time localization}

Assuming a terrain map is created with sufficiently crowd-sourced data, a vehicle that has the ability to reconstruct road profile in real-time and use this map to localize itself.

\begin{figure}
\centering
\includegraphics[width = 0.5\textwidth]{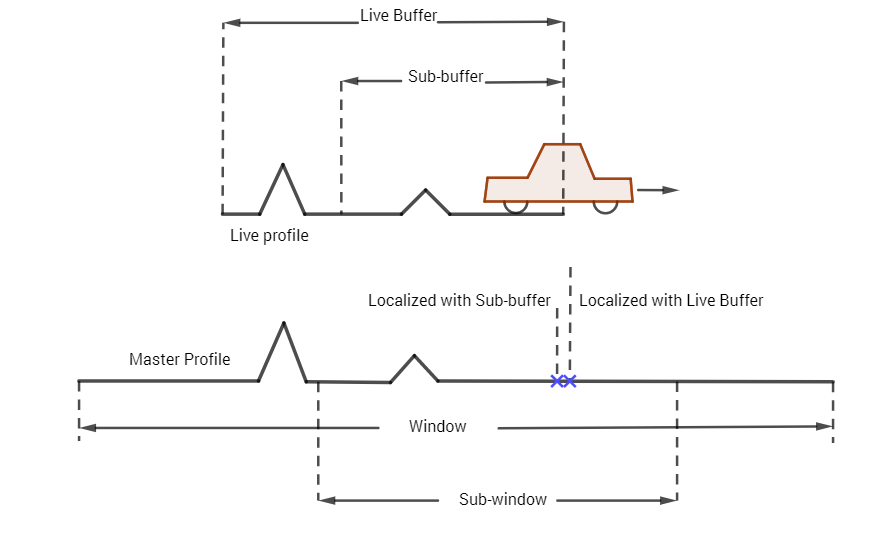}
\caption{Illustration of the localization algorithm. The on-board computer of the vehicle maintains a buffer, with a fixed length, of the most recently reconstructed road profile in distance. Based on any a priori knowledge of the vehicle's location, a window of master profile data is downloaded from the cloud. The vehicle then localizes itself by finding the relative location of its buffer within in the window. To attenuate the noise cause by distortion, a sub-window is imposed around the estimated location of the vehicle, and the algorithm tries to find a sub-buffer in the sub-window to refine the localization result.} 
\label{fig:xcorr}
\end{figure}

To be more specific, the on-board computer downloads the master pitch profile of a desired path from the cloud. Also, it takes the sensor data from the active suspension system, and applies Algorithm \ref{alg:profileconv} to create the profiles and maintain a buffer of the trailing  profile data. Then, the computer continuously matches its buffer within the master pitch profile to achieve localization. To avoid any distortion or distance discrepancy within the buffer data, the smaller buffer can be added to refine the localization results (Fig. \ref{fig:xcorr}).

% \subsection{Practical implementation}
% {\bf Something about the RoadMotion architecture?} 
% \begin{figure}
% \centering
% \includegraphics[width= 8cm, height = 7cm]{placeholder.png}
% \caption{RoadMotion architecture.}
% \end{figure}

\section{EXPERIMENTAL VALIDATION}
\subsection{Experimental setup}
\begin{figure}
\centering
\includegraphics[width= 7cm, height = 6cm]{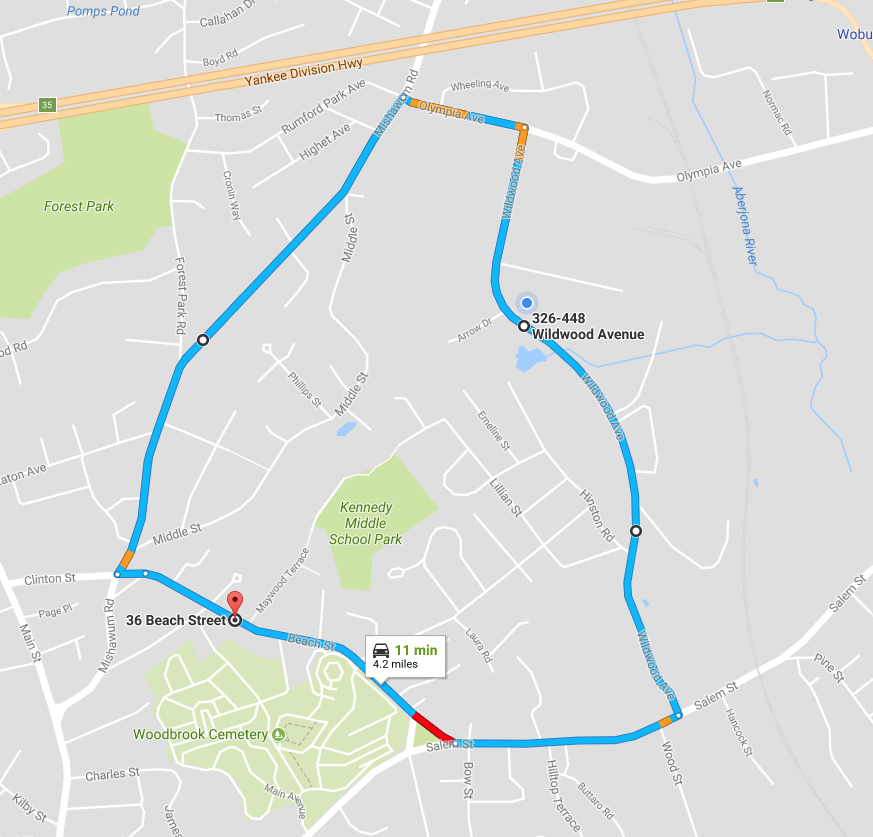}
\caption{Loop for experiments. This is a loop of local single lane road in Woburn, MA, USA.}
\label{fig:loop3}
\end{figure}

The experiments were conducted in a $4.2km$ loop in a residential area in Woburn, MA (Fig. \ref{fig:loop3}). First, we drove in this loop for three times to collect the profile data such that we can create an initial map. Then, we drove in the same loop to experiment with the real-time localization algorithm. 

\begin{figure}
\centering
\includegraphics[width= 9cm, height = 6cm]{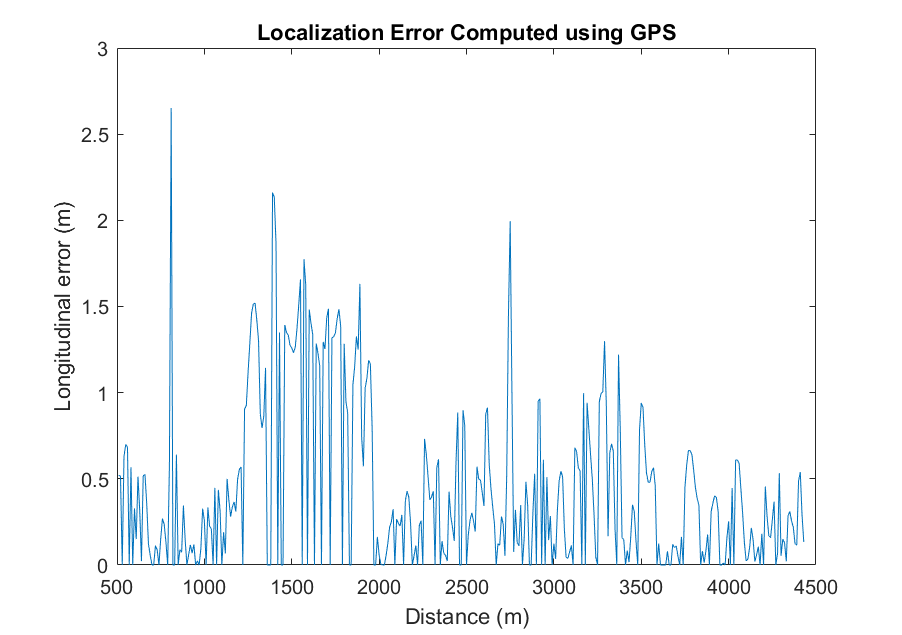}
\caption{Longitudinal error computed using on-board GPS measurements as the ground truth.}
\label{fig:err}
\end{figure}

\begin{figure}
\centering
\includegraphics[width= 8cm, height = 5.5cm]{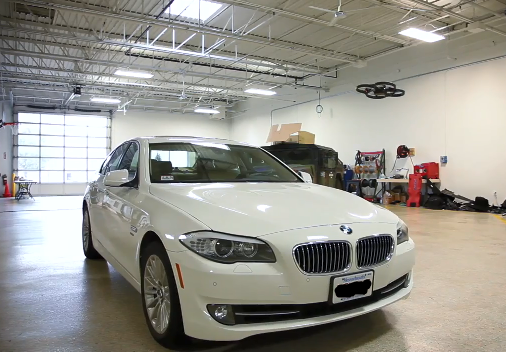}
\caption{Experimental vehicle (2012 BMW 535i)}
\label{fig:wallace}
\end{figure}

A 2012 BMW 535i was used in the experiment (Fig. \ref{fig:wallace}). It was equipped with a fully active suspension (FAS) system produced by ClearMotion, Inc. The FAS system includes four proprietary fast-response, compact, electrohydraulic acturators. They include a modified hydraulic damper integrated with a hydraulic motor-pump power unit, which enables the actuator to effectively control the motion of the vehicle body. The Genshock system also includes multiple sensors, as accelerometers, to measure each chassis corner's vertical acceleration.

%\begin{figure}
%\centering
%\includegraphics[width= 6cm, height = 8cm]{genshock.png}
%\caption{Active suspension system, developed by ClearMotion, Inc.}
%\label{fig:genshock}
%\end{figure}

\subsection{Profile reconstruction}
\begin{figure}
\centering
\includegraphics[width= 9cm, height = 6cm]{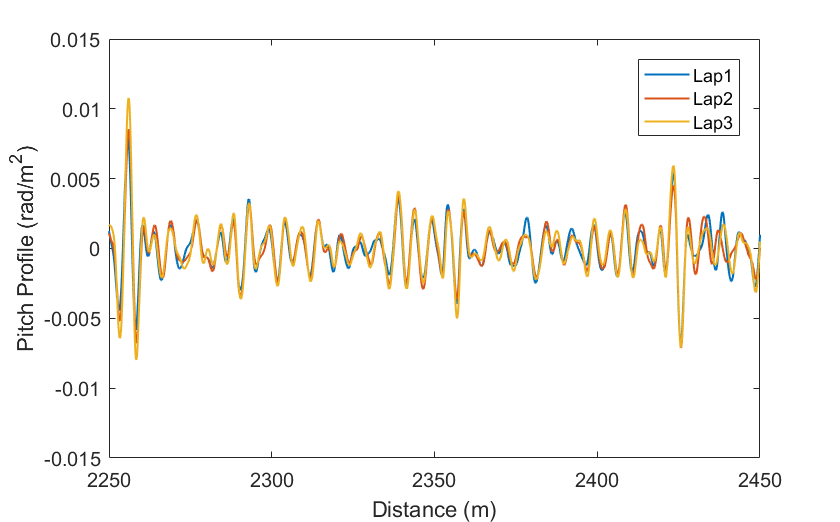}
\caption{Consistent pitch profile reconstruction}
\label{fig:pitch}
\end{figure}

To obtain consistent road profiles, we differentiate pitch angle with respect to distance for two times, and aligned the resultant profiles, as illustrated in Fig. \ref{fig:pitch}. It can be seen that the pitch profiles are highly consistent across three different laps. 

\subsection{Cross-correlation for localization}
\begin{figure}
\centering
\includegraphics[width= 9.25cm, height = 7cm]{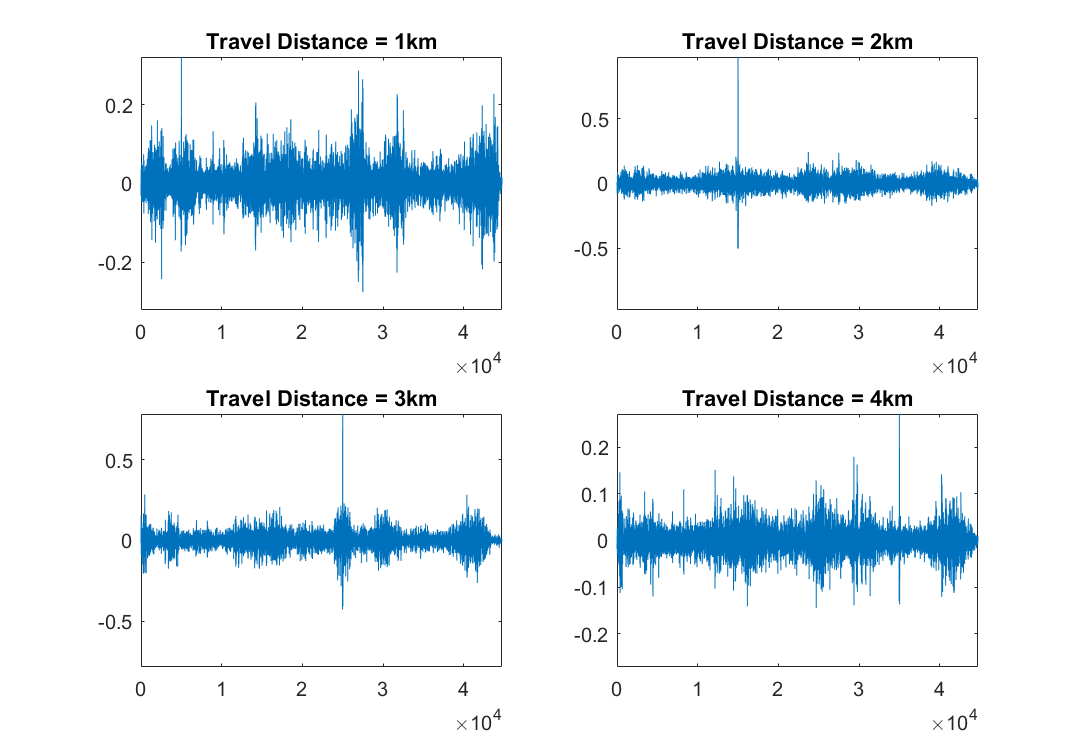}
\caption{Snapshots of cross-correlation at varies locations. The strongest correlation coincides with the location of the car's travel distance subtracted by the length of the buffer.}
\label{fig:result_plot}
\end{figure}

We perform continuous cross-correlation between the live buffer and the downloaded terrain map. Four snapshots of the cross-correlation at varies locations are shown in Fig. \ref{fig:result_plot}.
The strongest cross correlation gives the current location of the car relative to the map. By using the previously detected location of the vehicle as a priori knowledge, we  imposed a moving window of $1km$ on the downloaded profile data stream to significantly reduce chances of mismatching. When it was not possible to identify the strongest correlation point, we applied dead reckoning using only the vehicle speed to approximate the location and move the window. Once a clear spike is identified, the estimated location can be immediately corrected. In our experiments, more than $95\%$ of the time a strongest peak was clear (the second strongest peak was less than $60\%$ of the strongest peak).

\subsection{Error analysis}
We used the on-board GPS to compute the longitudinal error of the localization results. Error was computed at every $10m$ as shown in Fig. \ref{fig:err}.The error is less than $1m$ for more than $80\%$ of the time, less than $0.5m$ for more than $50\%$ of the time and $0.1m$ for more than $10\%$ of the time. Notice that the error is within GPS tolerance, indicating our accuracy is actually higher than computed based on GPS measurements.

\section{CONCLUSIONS AND FUTURE WORK}
In this paper, we have proposed a way to achieve localization using the sensor data from active suspension control systems. We have shown that it is practical to reconstruct road profiles and generate consistent a pitch profile that can be used for localization. A map creation and improvement approach has been presented to maintain a crowd-sourced terrain map. This method has been validated through experiments.

We believe the proposed methodology serves as a foundation for many related research directions. For example, it will be interesting to extend this approach to study localization with multiple lanes. Also, the pitch profile we generated can be combined with other sensor data such as yaw rate and steering angle, and terrain-related information such as special road features or extrema features \cite{vemulapalli2011pitch} to create a map with different layers. It may also work in combination with vision data \cite{gruyer2014map} to achieve both longitudinal and lateral localization. As the size of the terrain map increases, it is of importance to consider how to efficiently store the profile data or compress the data, e.g., using model-based methods \cite{laftchiev2013robust}. Finally, it is of great value to use the localization results as way to provide preview information \cite{roadSurface, chen2022vehicle, jiang2022vehicular} to active suspension systems for better ride improvement and motion-sickness mitigation \cite{ekchian2016high}.

\bibliographystyle{IEEEtran}
\bibliography{IEEEabrv,mybibfile}

\end{document}